\documentclass[lettersize,journal]{IEEEtran}
\usepackage{amsmath,amsfonts}
\usepackage{algorithmic}
\usepackage{array}
\usepackage{booktabs}
\usepackage[caption=false,font=normalsize,labelfont=sf,textfont=sf]{subfig}
\usepackage{textcomp}
\usepackage{stfloats}
\usepackage{url}
\usepackage{verbatim}
\usepackage{graphicx}
\def\BibTeX{{\rm B\kern-.05em{\sc i\kern-.025em b}\kern-.08em
    T\kern-.1667em\lower.7ex\hbox{E}\kern-.125emX}}
\usepackage{balance}
\usepackage{xcolor}
\usepackage{cite}

\begin{document}

\title{Appearance-Preserving Refinement of Generated 3D Assets for Monochromatic Fabrication}
\author{Chentao Shen*, Chen Jia*, Mingjie Huang, Zhuang Zhang, Haisen Zhao, Xiangru Huang
\thanks{\noindent*Chentao Shen and Chen Jia contributed equally to this work. Corresponding author: Xiangru Huang and Zhuang Zhang.

Chentao Shen is with Zhejiang University and with Westlake University (email: shenchentao@zju.edu.cn). Chen Jia and Zhuang Zhang are with Fudan University (email: cjia25@m.fudan.edu.cn, zhangzhuang@fudan.edu.cn). Mingjie Huang is with Hangzhou Dianzi University (email: 24030221@hdu.edu.cn). Haisen Zhao is with Shandong University (haisenzhao@sdu.edu.cn), Xiangru Huang is with Westlake University (email: huangxiangru@westlake.edu.cn)}}

\markboth{Arxiv Preprint}
{Geometric Appearance Refinement under Monochromatic Fabrication for 3D Generation}

\maketitle

\begin{abstract}

Recent advances in 3D mesh generation have enabled the creation of visually realistic assets. However, much of their visual fidelity is encoded in textures rather than geometry. When such assets are fabricated using monochromatic materials, texture information is largely lost, causing visually important details to disappear even when the original geometry is faithfully preserved.
A key challenge is that the geometric perturbations required to recover texture-dependent appearance cues often introduce sharp local features and high-frequency surface structures, which may increase stress concentration and fabrication risk. In this paper, we present \textbf{GenMF}, an appearance-oriented geometry refinement framework for monochromatic fabrication. GenMF transforms texture-dependent visual cues into geometry-induced shading effects and formulates geometry refinement as a balance between appearance preservation and fabrication-oriented robustness. To discourage structurally and narrow the gap between simulation and physical manufacturing, we further introduce a differentiable stress-aware regularization based on a learned thermal-stress predictor.
Experimental results demonstrate that GenMF significantly improves appearance preservation under monochromatic rendering while reducing stress concentration under a consistent thermo-mechanical simulation setting. Physical 3D printing examples further show that the refined geometries preserve more recognizable visual details while remaining suitable for fabrication. These results suggest that appearance-aware geometry refinement provides an effective bridge between generated 3D assets and fabrication-ready monochromatic objects.

\end{abstract}



\section{Introduction}
\label{sec:intro}

Recent advances in image-to-3D and text-to-3D generation such as Trellis\cite{trellis} have enabled the rapid creation of high-quality 3D assets. Modern generation systems can produce visually realistic objects with rich appearance details, making them increasingly useful for content creation, digital entertainment, and virtual environments. As a result, generated 3D assets are becoming an important source of 3D content for both professional and consumer applications.

\begin{figure}
    \centering
    \includegraphics[width=0.92\linewidth]{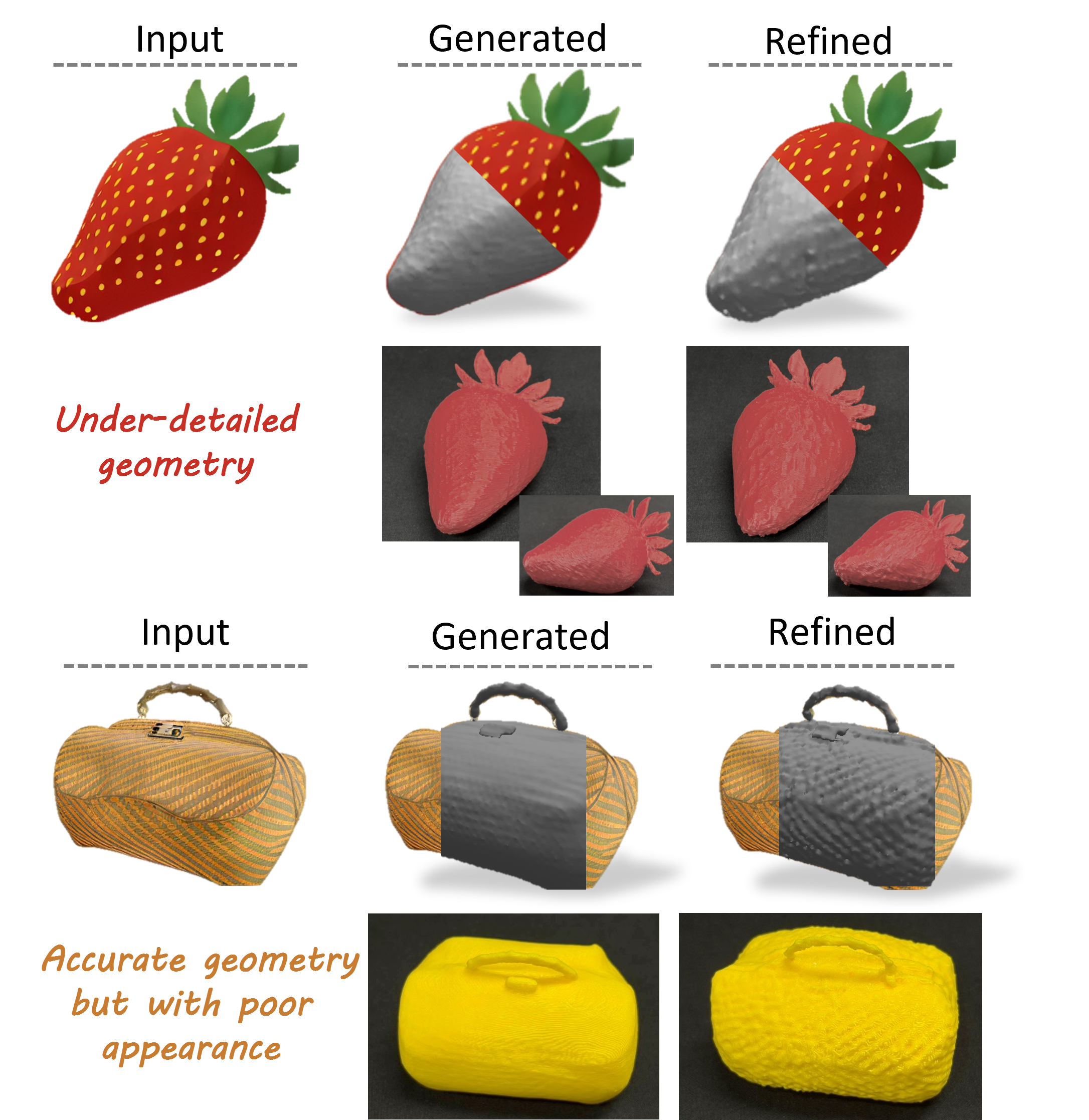}
    \caption{Texture-dependent details such as veins, seeds, semantic boundaries, and surface patterns are visible in textured digital assets but disappear after monochromatic fabrication. This motivates geometry-based appearance encoding.}
    \label{fig:tu1}
\end{figure}

Despite their visual quality, most generated assets rely heavily on textures to represent fine-scale appearance details. When these assets are transferred from virtual environments to physical fabrication, such appearance cues can no longer be preserved directly. 

Specifically, most conventional manufacturing processes and additive manufacturing pipelines remain dominated by single-material fabrication\cite{NAZIR2023111661}. Although multi-material and full-color printing technologies have attracted increasing attention, they typically require specialized hardware and introduce substantial fabrication complexity. Consequently, many practical fabrication systems are still restricted to a single or a limited set of discrete color/material combinations, making geometric cues a primary carrier of visual appearance\cite{fabbaloo_color2024}. In these processes, texture information is largely discarded, and the \textbf{final visual appearance is primarily determined by geometry-induced shading}. 

As a result, preserving the original geometry does not necessarily preserve the perceived appearance of the object after fabrication. For example, as shown in Fig.~\ref{fig:tu1}, texture-dependent details, such as color boundaries, surface patterns, veins, seeds, and semantic markings, may become indistinguishable once the object is fabricated with a single material. After 3D Printing, these texture-dependent details disappear. Therefore, the desired geometry for fabrication is not necessarily the one that best preserves the original shape, but the one that reproduces the perceived appearance under monochromatic rendering like geometry texture~\cite{zhou2020deepgeometrictexture}.

However, \textbf{appearance preservation and fabrication-oriented robustness are inherently coupled}. The fine-scale geometric perturbations required to recover texture-dependent appearance cues often introduce sharp local features and high-frequency surface structures. While such modifications may improve appearance in digital renderings, they can also introduce stress concentrations and reduce structural robustness, making fabricated objects more susceptible to deformation or structural failure under fabrication\cite{zheng2024quantitative}. Fully modeling structural robustness in fabrication requires material-specific parameters, process settings, support conditions, and external loads, which are difficult to generalize across diverse generated objects. In this work, we use thermal stress under a standardized thermo-mechanical simulation setting as a differentiable proxy for fabrication-oriented structural robustness. This proxy does not cover all possible manufacturing failures, but it provides useful guidance for discouraging geometrically unstable refinements.

To address this challenge, we propose \textbf{GenMF}, an appearance-oriented geometry refinement framework for  monochromatic fabrication like 3D Printing. Our framework is implemented as a plug-and-play post-refinement module that can be applied to existing 3D generation pipelines (e.g., SDF-based InstantMesh, SDF\&deform-based Stable3DGen, Trellis). It refines the geometry so that visually important texture cues are expressed through geometry-induced shading. The optimization is guided by differentiable monochromatic rendering losses that compare the refined geometry with reference appearance cues in grayscale space. To improve structural robustness and decrease the fabrication-induced deformation to narrow the gap between simulation and physical manufacturing, we further introduce a differentiable stress-aware regularization term based on a learned thermal-stress predictor. 

Based on the observation that appearance recovery and stress concentration originate from the same geometry modifications, GenMF formulates geometry refinement as a trade-off between perceptual detail reconstruction and fabrication-oriented robustness. Importantly, GenMF is model-agnostic and can be applied to textured assets generated by different image-to-3D and text-to-3D pipelines without modifying the underlying generation models.

The main contributions of this paper are as follows:

\begin{itemize}
\item We identify and address the trade-off between appearance preservation and fabrication-oriented robustness in monochromatic fabrication of generated 3D assets, formulating geometry refinement as a joint optimization problem over these competing objectives.
\item We introduce a differentiable stress-aware robustness prior that suppresses high-frequency surface perturbations associated with stress concentration while preserving appearance-relevant geometric details.
\item We demonstrate through quantitative evaluation, ablation studies, FEM simulation, and physical fabrication that GenMF improves monochromatic appearance preservation while reducing stress concentration.
\end{itemize}
\section{Related Work}




\textbf{3D Generation.} Recent advances in 3D generation have enabled the creation of high-quality 3D assets from text, single images, or sparse multi-view observations. Existing methods are commonly based on explicit mesh representations~\cite{gao2022get3d, Magic3D, xu2024instantmesh, trellis,liu2024meshformer, tsalicoglou2024textmesh, lin2025partcrafter}, implicit neural fields~\cite{poole2023dreamfusion, ICLR2024mvdream, long2024wonder3d, ProlificDreamer}, or Gaussian-based representations~\cite{yi2024gaussiandreamer, tang2023dreamgaussian, tang2024lgm, long2024wonder3d, tang2023makeit3d, wu2024directd, wu2025directds, lan2025ln3diff++, Magic123}. While these methods achieve impressive visual realism, much of their appearance fidelity remains encoded in textures or view-dependent effects, making them less suitable for monochromatic fabrication scenarios. 

\textbf{Geometry Refinement for 3D Generation.} Recent advances in 3D generation have increasingly explored the recovery of geometric details from appearance supervision. DreamFusion\cite{poole2023dreamfusion} refine a NeRF representation using Score Distillation Sampling (SDS), while Magic3D\cite{Magic3D} refines extracted meshes with high-resolution diffusion guidance. However, both methods primarily rely on RGB-rendered supervision, causing appearance details to be absorbed into textures rather than geometry. Fantasia3D~\cite{chen2023fantasia3d} disentangles geometry and appearance by optimizing DMTet-based SDFs using diffusion supervision on rendered normal maps, encouraging high-frequency appearance features to be encoded as geometry rather than texture. Fancy123~\cite{yu2025fancy123} further improves single-image 3D reconstruction through multi-view diffusion guidance, jointly optimizing geometry and appearance under stronger view-consistency constraints. More recently, generative mesh models such as Craftsman3D~\cite{li2025craftsman3d} and Hi3DGen~\cite{ye2025hi3dgen} learn high-frequency geometric priors from large-scale mesh datasets. Although they do not perform explicit geometry refinement at inference time, their training objectives implicitly capture fine-scale geometric patterns and demonstrate the effectiveness of data-driven geometric detail modeling. However, these methods primarily aim to \textbf{recover geometrically accurate or visually realistic surfaces}. In contrast, our goal is not to recover the most faithful geometry, but to intentionally reshape geometry so that texture-dependent appearance cues remain visible without color.

\textbf{Appearance-Preserving Geometry Refinement.} Enhancing visual appearance through geometric modification has long been studied in computer graphics. A common strategy is to encode appearance details using displacement or displacement-like representations \cite{HFNEUS,D2IMNET,yifan2021geometry}, where high-frequency surface structures are modeled as offsets from a coarse mesh or implicit surface. While effective for recovering fine geometric details, these methods are primarily designed for faithful surface reconstruction and remain limited in representing complex geometric patterns and appearance-driven structures \cite{zhou2020deepgeometrictexture}. Another line of research represents visual textures directly as geometry through geometric texture synthesis \cite{zhou2020deepgeometrictexture,zhoukun2006}, enabling the generation of rich surface details without relying on texture maps. However, these methods typically require accurate initial geometry and predefined geometric texture examples, making them difficult to generalize to objects containing multiple heterogeneous appearance patterns.

\textbf{Structural Robustness Refinement.} Guaranteeing the physical viability and structural soundness of synthesized shapes represents a core endeavor at the intersection of geometry processing and computational fabrication. Early frameworks primarily rely on geometric modifications, such as hollowing and local thickening~\cite{ssr1,ssr2} or internal topology tuning~\cite{ssr3,ssr4}, to remedy localized weak points against worst-case mechanical loads under isothermal conditions. To alleviate the heavy computational overhead of these static optimization pipelines, recent methods focus on interactive, physics-guided shape design by accelerating elastostatic analysis and inverse elastic optimization~\cite{ssr5} to deliver instantaneous stress feedback. More recently, the field has expanded toward multi-physics coupling, integrating complex constraints such as combined aerodynamic-mechanical loads, transient temperature fields, and heterogeneous multi-material layouts to mitigate severe thermoelastic stresses~\cite{ssr6,ssr7,ssr8,ssr9}. However, these approaches fundamentally prioritize engineering load satisfaction over visual fidelity, often sacrificing high-frequency geometric details to ensure physical viability. Unlike these methods, we do not aim at load-optimal engineering design; instead, we use stress feedback as a regularizer to prevent appearance-driven refinements from creating structurally risky details.

\textbf{Data-driven Stress Prediction.} Deep learning has become an efficient alternative to finite element method (FEM) simulations for stress field prediction. Early approaches employed convolutional neural networks (CNNs) and generative adversarial networks (GANs), such as U-Net~\cite{s1,s2,s3,s4} and conditional GANs~\cite{s5,s6,s7,s8}, to map geometries or microstructures to stress distributions. Recent methods incorporate Transformer-based architectures, e.g., 3D TransU-Net~\cite{s9}, to model volumetric stress fields under varying loading conditions. For irregular domains, graph neural networks (GNNs)~\cite{s10,s11,s12,s13} predict stress fields on mesh structures, while physics-informed neural networks (PINNs)~\cite{s14,s15,s16} embed governing equations to reduce reliance on labeled data. More recently, neural operator frameworks, including Fourier Neural Operators~\cite{s17,s18} and DeepONet~\cite{s19,s20,s21}, learn mappings between function spaces and show strong generalization across unseen geometries and boundary conditions.

\section{Methodology}
\label{sec:method}

\subsection{Overview}

\begin{figure*}
    \centering
    \includegraphics[width=0.9\linewidth]{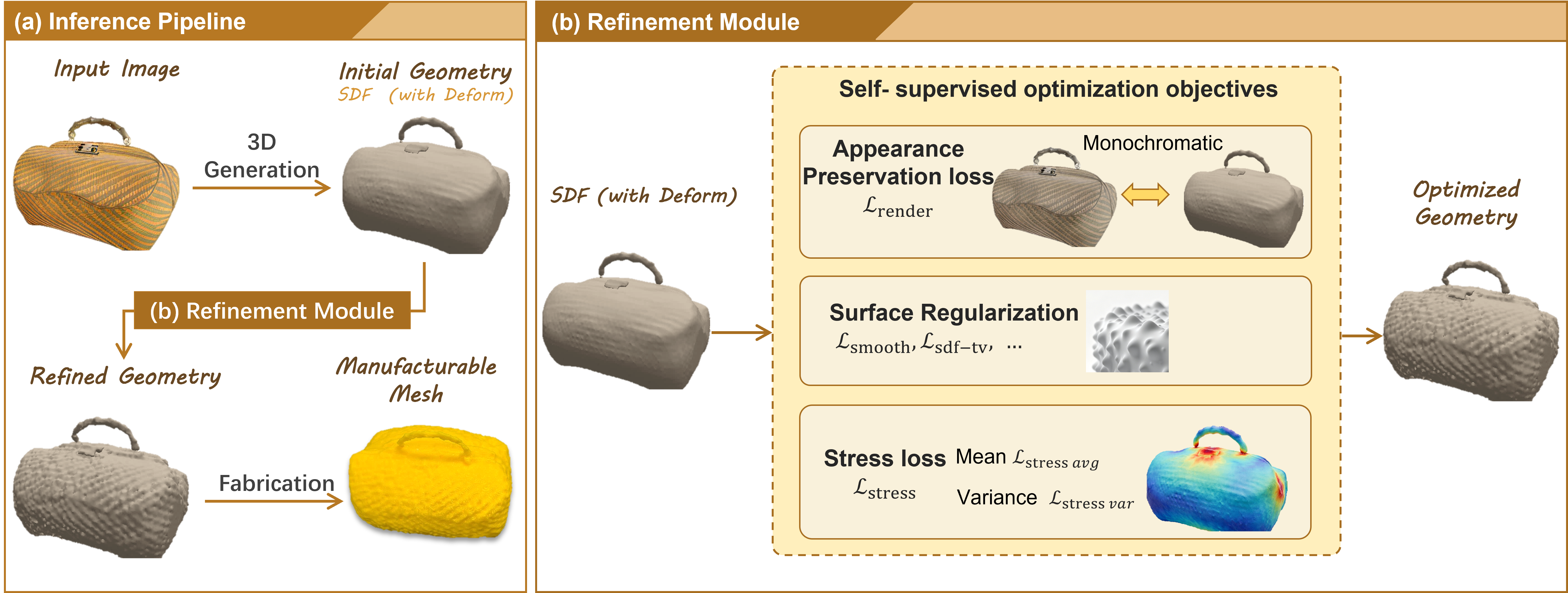}
    \caption{Overview of GenMF. Given a textured 3D asset generated by an existing image-to-3D model, GenMF refines its geometry through monochromatic appearance supervision, surface regularization, and stress-aware feedback, producing a fabrication-oriented geometry with improved texture-free appearance.}
    \label{fig:mainpipeline}
\end{figure*}


Given a 3D asset generated by existing methods (e.g., InstantMesh, Hi3DGen, or Trellis), our goal is not to faithfully preserve the original geometry, but to produce a structurally stable geometry that retains its visual appearance after monochromatic fabrication.

We formulate this task as an appearance-oriented geometry optimization problem. In monochromatic fabrication, texture information is largely unavailable, causing many visually important details to disappear even when the underlying geometry remains unchanged. Consequently, faithful geometry reconstruction alone does not necessarily lead to faithful appearance preservation after fabrication. To address this challenge, we seek to transform texture-dependent visual cues into geometry-induced shading effects while maintaining structural robustness.

Specifically, our optimization jointly considers three objectives:


\begin{itemize}

\item \textbf{Monochromatic Appearance preservation}: preserving the perceived visual appearance of the input asset under monochromatic fabrication by encoding texture-dependent details into geometry;
\item \textbf{Structural robustness}: suppressing stress concentration and reducing fabrication-induced deformation;
\item \textbf{Geometric regularity}: enforcing surface smoothness and preventing optimization artifacts.
\end{itemize}

These objectives are inherently coupled and often conflicting. For example, enhancing appearance through local geometric perturbations may improve visual detail reproduction, but can simultaneously introduce stress concentration and structural fragility. Conversely, excessive regularization may suppress visually important features. Our framework explicitly balances these competing objectives within a unified optimization process.

To this end, we propose a refinement framework consisting of (1) a reference-guided differentiable appearance-driven optimization module and (2) a differentiable stress prediction network that provides structural feedback during optimization. 

\subsection{Structural robustness}

To improve the structural robustness of generated meshes, we introduce stress-aware optimization into the refinement process.

Directly modeling physical stress typically requires task-specific boundary conditions and external forces, which are difficult to generalize across diverse generated shapes. Instead, we adopt \textbf{thermal stress} under a standardized thermo-mechanical setup as a differentiable proxy that correlates with thin structures, sharp features, and local stress concentration.

We employ a neural stress prediction network (detailed in Sec. \ref{sec:stresspre}) to estimate the stress field from the implicit geometry. This enables a fully differentiable mapping from shape to structural response, allowing gradients to propagate back to the geometry during optimization.

Based on the predicted stress field, we define two complementary objectives: minimizing the overall stress magnitude ($\mathcal{L}_{\text{stress avg}} = \mathrm{E}(\sigma)$) and encouraging uniform stress distribution ($\mathcal{L}_{\text{stress var}} = \text{Var}(\sigma)$). The former improves global stability, while the latter reduces local stress concentration, which is a common cause of deformation in manufacturing.

This formulation provides an efficient and generalizable approximation of structural optimization, enabling geometry refinement guided by physical considerations.



\subsection{Monochromatic Appearance Preservation}


Another challenge in manufacturable generation is that existing methods encode fine details primarily in textures, which are unavailable in monochromatic fabrication. To address this, we aim to transfer texture-dependent appearance into geometry.

\begin{figure}
    \centering
    \includegraphics[width=0.96\linewidth]{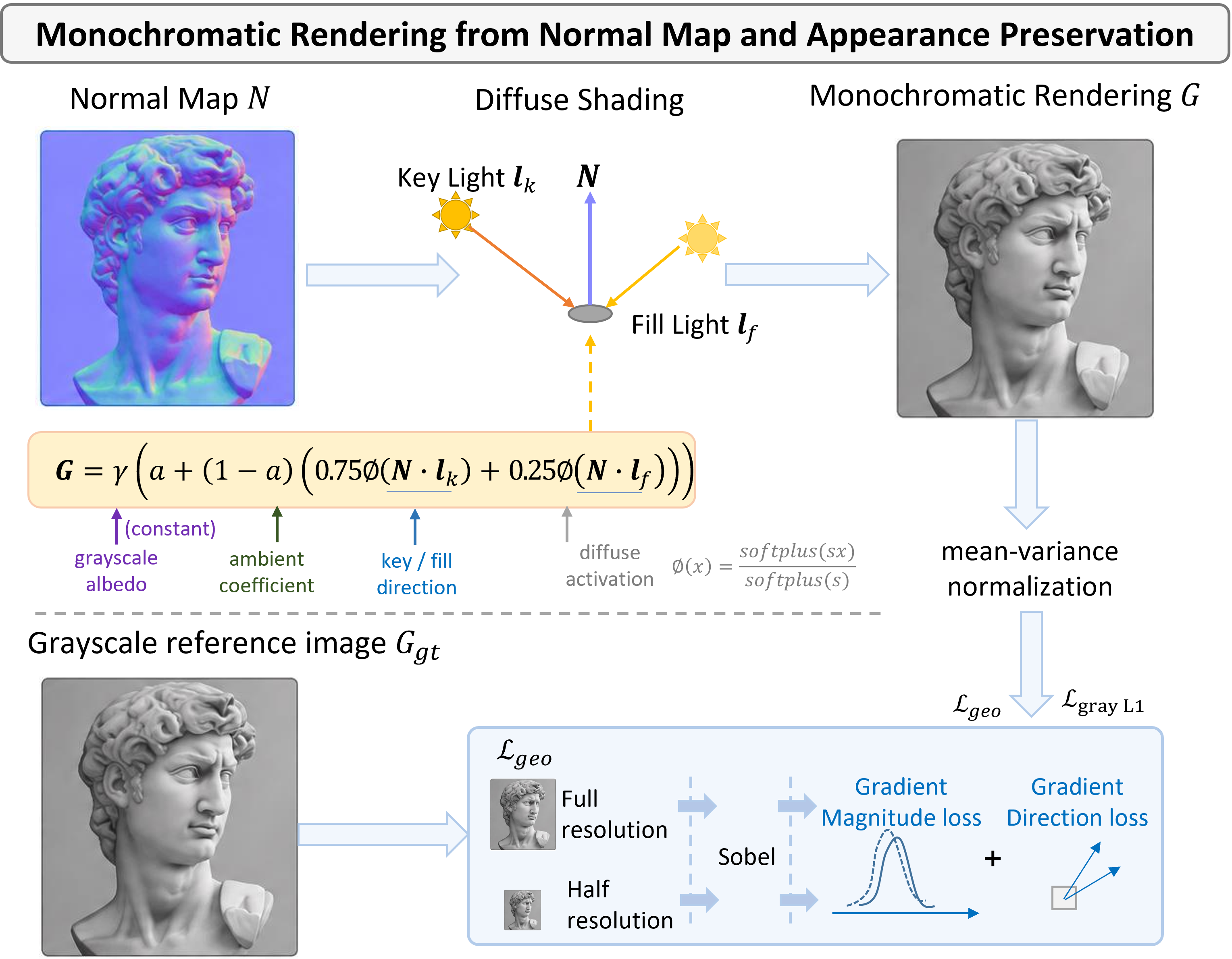}
    \caption{The rendering pipeline and the appearance supervision}
    \label{fig:apppre}
\end{figure}

Rather than directly supervising texture maps (which introduces non-differentiability and instability), we operate in the \textbf{non-texture (monochromatic)} rendering space. This is motivated by the observation that geometric details are primarily reflected through shading variations in grayscale images, as shown in Fig.~\ref{fig:apppre}.

For efficiency, we directly calculate the monochromatic rendering image $G\in \mathbb{R}^{H \times W}$ by normal map $N\in \mathbb{R}^{H \times W \times 3}$ based on a fixed physically-inspired shading model. 

The normal vectors are first normalized into unit surface normals, and diffuse shading is computed under two predefined directional light sources: a frontal key light and a secondary fill light. To obtain smooth and differentiable shading responses, we replace the standard Lambertian activation with a softplus-based approximation.The final monochrome rendering is defined as:
\begin{equation}
G = \gamma \left(
a + (1-a)\left(
0.75\,\phi(N \cdot l_k) +
0.25\,\phi(N \cdot l_f)
\right)
\right),
\end{equation}
where $\gamma$ denotes the grayscale albedo (a constant for whole object), $a$ is the ambient lighting coefficient, $l_k$ and $l_f$ represent the key and fill light directions, respectively, and $\phi(\cdot)$ is a smooth diffuse activation implemented using a normalized softplus function with $s$ controlling the lighting sharpness.

\begin{equation}
\phi(x) = {\mathrm{softplus}(s x)}/{\mathrm{softplus}(s)}
\end{equation}

Before computing the geometric feature loss, we normalize the monochrome rendering $G$ to match the intensity statistics of the reference grayscale image $G_{\mathrm{gt}}$. Instead of using a standard grayscale conversion, we define $G_{\mathrm{gt}}$ as the value channel in HSV space, which better captures illumination and shading variations while reducing the influence of surface color.
This alignment removes global brightness and contrast differences caused by illumination variations, allowing the supervision to focus on structural shading patterns. Specifically, we perform mean-variance normalization:
\begin{equation}
\hat{G}=\frac{G - \mu_G}{\sigma_G}\sigma_{G_{\mathrm{gt}}}+\mu_{G_{\mathrm{gt}}},
\end{equation}
where $\mu_G, \sigma_G$ and $\mu_{G_{\mathrm{gt}}}, \sigma_{G_{\mathrm{gt}}}$ denote the mean and standard deviation of the rendered image and target grayscale image, respectively.
The aligned rendering $\hat{G}$ is then used for subsequent gradient-based appearance supervision.

Consequently, we first introduce a set of monochromatic rendered losses  $\mathcal{L}_{\text{gray L1}}$ and $\mathcal{L}_{\text{gray Lpips}}$ between the monochromatic rendered output $G$ and the reference grayscale-image $G_{gt}$ (the gray-scale image of input or Zero123). These constraints encourage the geometry to reproduce visual details through shading rather than texture.

Additionally, we introduce a geometric feature loss $\mathcal{L}_{\mathrm{geo}}$ defined between the monochrome rendering $G$ and the reference grayscale image $G_{\mathrm{gt}}$. It compares their structural shading patterns using multi-scale Sobel gradient features, which better capture geometry-induced brightness variations. For each image, horizontal and vertical gradients are extracted using Sobel operators:
\begin{equation}
\nabla \hat{G} = (G_x, G_y), \qquad
\nabla G_{\mathrm{gt}} = (G_x^{\mathrm{gt}}, G_y^{\mathrm{gt}}).
\end{equation}

We then compute both gradient magnitude consistency and gradient direction consistency. 
The magnitude loss is defined as:
\begin{equation}
\mathcal{L}_{\mathrm{mag}}=\left\|\sqrt{G_x^2 + G_y^2}-
\sqrt{(G_x^{\mathrm{gt}})^2 + (G_y^{\mathrm{gt}})^2}\right\|_1,
\end{equation}
while the directional loss is formulated using cosine similarity:
\begin{equation}
\mathcal{L}_{\mathrm{dir}}=1 -
(\frac{G_x G_x^{\mathrm{gt}} + G_y G_y^{\mathrm{gt}}}{\|\nabla G\| \, \|\nabla G_{\mathrm{gt}}\|})^2.
\end{equation}

The final geometric feature loss combines both terms:
\begin{equation}
\mathcal{L}_{\mathrm{geo-}}=\mathcal{L}_{\mathrm{mag}}+0.5\,\mathcal{L}_{\mathrm{dir}}.
\end{equation}

To improve robustness to scale variation, we further compute the loss at both the original resolution and a downsampled resolution, and combine them as:
\begin{equation}
\mathcal{L}_{\mathrm{geo}}=0.6\,\mathcal{L}_{\mathrm{full}}+0.4\,\mathcal{L}_{\mathrm{half}}.
\end{equation}



\subsection{Geometry regularization}

To ensure stable optimization and high-quality surfaces, we introduce additional regularization terms on the geometry. These regularizations serve two purposes: (1) suppressing high-frequency artifacts introduced during optimization, and (2) maintaining structural coherence of the mesh. 

Specifically, we enforce smoothness through depth ($\mathcal{L}_{\text{depth sm}}$), normal ($\mathcal{L}_{\text{norm sm}}$), and SDF-based regularization ($\mathcal{L}_{\text{sdftv}}$), which separate noise from underlying geometric structure and apply constraints on high-frequency components. where $\nabla^2=[\nabla_x^2,\nabla_y^2]^T$ denotes the second-order spatial derivative operator.



\begin{equation}
\begin{aligned}
    \mathcal{L}_{\text{depth sm}} = \mathbb{E}\left[\left\|\nabla^2 \left(D - \text{blur}(D)\right)\right\|_1\right]\\
    \mathcal{L}_{\text{norm sm}} = \mathbb{E}\left[\left\|\nabla^2 \left(N - \text{blur}(N)\right)\right\|_2\right]
    \end{aligned}
\end{equation}

\begin{equation}
    \mathcal{L}_{\text{sdf-tv}} = \sum \mathbb{E}\left[\left|\nabla \Phi\right|\right]
\end{equation}

We further introduce mesh-level constraints to control complexity, including surface and weight regularization for FlexiCubes ($\mathcal{L}_{\text{reg}}$ in InstantMesh\cite{xu2024instantmesh}).

Finally, to preserve the overall structure of the initial reconstruction, we adopt a teacher–student supervision strategy using depth and normal consistency. This prevents the optimization from deviating excessively from the original geometry while allowing local refinement. We define $N_t,D_t$ as the normal and the depth map of the initial reconstruction.


\begin{equation}
    \begin{aligned}
\mathcal{L}_{\text{teac-depth}} &= \mathbb{E}\left| D(\mathbf{x}) - D_t(\mathbf{x}) \right|, \\
\mathcal{L}_{\text{teac-norm}} &= 1 - \mathbb{E}\left| N(\mathbf{x}) \cdot N_t(\mathbf{x}) \right|,
\end{aligned}
\end{equation}

\subsection{Optimization-Guided Geometry Refinement }

The final objective combines all components into a unified multi-objective optimization. For SDF-based generators, we optimize the deformable SDF representation before mesh extraction; For SDF and Deform-based generators, we optimize both them with greater weight of SDF and a low but increasing weight of Deform.  The entire pipeline is end-to-end differentiable and can be seamlessly integrated as a plug-and-play refinement stage for existing 3D generation methods.

\begin{equation}
    \mathcal{L}_{\text{total}} = \sum_{name\in\{\text{geo},\text{stress},...\}} \lambda_{name}\mathcal{L}_{name}
\end{equation}

We adopt a staged optimization strategy to stabilize training. In the first stage, we focus on appearance–geometry decoupling by disabling stress-related losses, allowing the model to establish geometry-driven appearance. In the second stage, we enable stress optimization to further refine the geometry toward structurally stable configurations.

This staged design reflects the inherent trade-off between appearance and physical constraints, and helps avoid suboptimal solutions during joint optimization.


\section{Stress Prediction}
\label{sec:stresspre}

\begin{figure}
    \centering
    \includegraphics[width=0.98\linewidth]{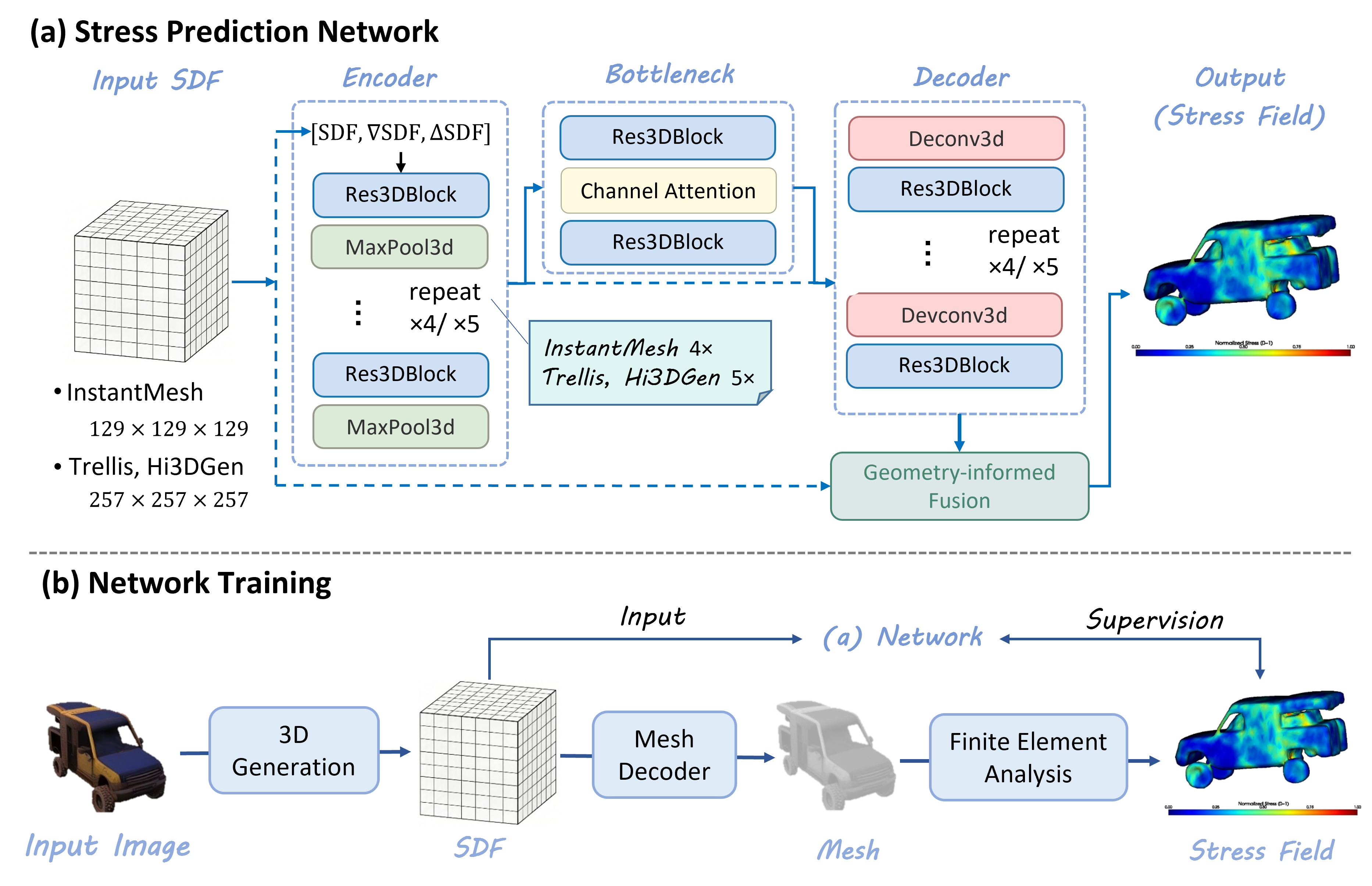}
    \caption{Stress Prediction Network and its training pipeline}
    \label{fig:placeholder}
\end{figure}

\subsection{Network Design}

This section details the design of the stress prediction module, which is engineered to map latent geometric representations into continuous 3D stress fields by extracting critical structural features from implicit geometry. By establishing a differentiable mapping from geometry to mechanical response, the network allows stress-related loss to propagate gradients directly back to the geometric parameters, enabling end-to-end structural optimization of desired objects.

\textbf{Geometric Feature Augmentation.} Mechanical stress distributions are highly sensitive to local surface orientations and curvatures. To provide the network with an explicit inductive bias for physical-informed learning, we implement a Geometric Gradient Layer  that computes first-order and second-order spatial features. Given an input SDF $\Phi \in \mathbb{R}^{D \times H \times W}$,  we compute the spatial gradients $\nabla \Phi$ and the Laplacian $\Delta \Phi \in \mathbb{R}^{1\times D \times H \times W}$ using finite difference operators. The gradient of the SDF represents the normal direction of the iso-surface, while the Laplacian $\Delta \Phi$ captures the divergence of the gradient field, serving as a proxy for mean curvature. To ensure numerical stability, we normalize the gradients to obtain the normalized gradient $\mathbf{n}\in \mathbb{R}^{3\times D \times H \times W}$.



\begin{equation}
    \mathbf{n}=\frac{\nabla \Phi}{\|\nabla \Phi\|},~~~ \Delta \Phi = \sum_{i \in \{x,y,z\}} \frac{\partial^2 \Phi}{\partial i^2}
\end{equation}
The augmented input feature map $\mathbf{X}_{in}$ is formed by concatenating the raw SDF, the normalized gradients, and the Laplacian:

\begin{equation}
    \mathbf{X}_{in} = \mathrm{Concat}(\Phi, \mathbf{n}, \Delta\Phi) \in \mathbb{R}^{5\times D\times H\times W}
\end{equation}
This 5-channel representation explicitly provides the network with second-order geometric information, significantly enhancing the model’s ability to capture stress concentrations near thin-walled boundaries and sharp edges.

\textbf{Hierarchical Residual Architecture.} The core prediction module utilizes a symmetric 3D Encoder-Decoder structure designed to handle high-resolution volumetric data ($129^3$). Each scale level utilizes Res3DBlock modules, which consist of two $3 \times 3 \times 3$ convolutional layers with Instance Normalization and ReLU activation. The residual connection is defined as:

\begin{equation}
    \mathbf{x}_{l+1} = \text{ReLU}(\mathcal{F}(\mathbf{x}_l, \{\mathcal{W}_l\}) + \text{Identity}(\mathbf{x}_l))
\end{equation}
The encoder employs four stages of max-pooling, reducing the spatial resolution from $129^3$ to $8^3$ while progressively increasing the channel capacity from $C_{base}=56$ to $8C_{base}$. This hierarchical downsampling allows the network to aggregate global structural stiffness information across an expanding receptive field.

\textbf{Global Context and Feature Recalibration.} Stress distribution is a global physical phenomenon where the response at any point depends on the overall topological connectivity and boundary conditions of the structure. To address this, we introduce two enhancement mechanisms in the network bottleneck. A Squeeze-and-Excitation (SE) module adaptively recalibrates channel-wise feature responses by modeling inter-dependencies between channels, emphasizing features most relevant to the specific mechanical topology. Moreover, we extract a global descriptor $\mathbf{z}_g$ by adaptive average pooling of the bottleneck and fuse it back into the spatial feature map using residual addition:

\begin{equation}
    \mathbf{z}_{fused} = \mathbf{z}_{bottleneck} + \mathrm{AvgPool}(\mathbf{z}_{bottleneck})
\end{equation}
This ensures that local stress predictions are grounded in the global context of the object’s mass distribution and macro-scale proportions.

\subsection{Data Generation \& Stress Network Training}

\textbf{Boundary-Consistent Fusion and Loss.} The decoder recovers spatial resolution via transpose convolutions, supplemented by skip connections that reintroduce high-frequency geometric details from the encoder. In the final stage, we implement a Geometry-Informed Fusion layer that concatenates the original SDF $\Phi$ with the decoded features before the final $1 \times 1$ convolution:

\begin{equation}
    \hat{\sigma} = \text{Conv}_{out}(\text{Conv}_{fusion}(\text{Concat}(\mathbf{X}_{dec1}, \Phi)))
\end{equation}
The “re-injection” of the SDF ensures that the predicted stress field $\hat{\sigma}$ is strictly aligned with the volumetric boundary of the object. For Training, we account for the wide dynamic range of mechanical stress by applying a non-linear mapping to the target values: 

\begin{equation}
    \tilde{\sigma}_{gt} = \log\left(1 + \frac{\sigma_{gt}}{k}\right)
\end{equation}
where $k=3.0$. The network is trained using Mean Squared Error (MSE) within the valid material domain ($\Phi \le 0$), ensuring balanced precision across both smooth low-stress regions and critical stress concentration zones.

\textbf{Data Preparation.} Our dataset is curated through a multi-stage geometric extraction process, leveraging InstantMesh to generate both implicit and explicit representations from 2D image inputs. (a) Implicit Feature Extraction: For each input image from the Arb-objaverse\cite{li2025idarb} dataset, we utilize InstantMesh to synthesize a set of tri-planes and a high-resolution SDF grid. Volumetric SDF serves as the primary input $\Phi$ for our stress prediction network, as detailed in 4.1. (b) Explicit Geometry Generation: Simultaneously, InstantMesh generates a raw 3D mesh. To ensure the mesh is suitable for physical simulation, we perform surface repair and mesh simplification to eliminate topological artifacts. The refined geometry is exported in STL format. These explicit models serve as the geometric foundation for subsequent Finite Element Analysis (FEA) to derive the ground truth stress distributions.

\textbf{Finite Element Analysis.} The ground truth stress ($\sigma_{gt}$) and displacement fields are obtained through a rigorous multiphysics simulation pipeline. To handle the complexity of large-scale manufacturing models, we implement a complete simulation workflow. (a) Geometric Reconstruction: We utilize HyperMesh (version: 2025) to perform 3D facet reconstruction on the initial STL files. This step ensures high-quality mesh connectivity and surface integrity, producing a BDF (Bulk Data Format) mesh suitable for volumetric solid modeling. (b) Multiphysics Coupling: BDF files are imported into COMSOL Multiphysics (version: 6.1) to construct solid entities. Polylactic Acid (PLA) is specified as the constituent material, reflecting its prevalence in consumer-grade 3D printing. We establish a transient study coupling Heat Transfer in Solids with Solid Mechanics. The simulation accounts for thermal expansion and internal stress development by defining fixed boundary constraints and a time-variant ambient temperature function $T_{amb}(t)$. The resulting transient thermal stress and displacement distributions are recorded as the ground truth. (c) Automated Large-Scale Synthesis: To address the computational bottleneck of generating thousands of simulations, we develop an integrated automation script using MATLAB (version: R2025b). This script handles data communication between HyperMesh and COMSOL, enabling one-click batch processing from raw STLs to final stress fields. The pipeline significantly reduces the temporal cost and manual complexity of constructing a large-scale, physics-aligned dataset for 3D deep learning.

\section{Experiments}
\label{sec:experiments}

\subsection{Experimental Setup} 
\label{sec:sec51}
We evaluate our method from three perspectives: (1) appearance-to-geometry transfer, (2) structural robustness, and (3) the trade-off between visual fidelity and structural robustness. 

\begin{figure*}
    \centering
    \includegraphics[width=0.85\linewidth]{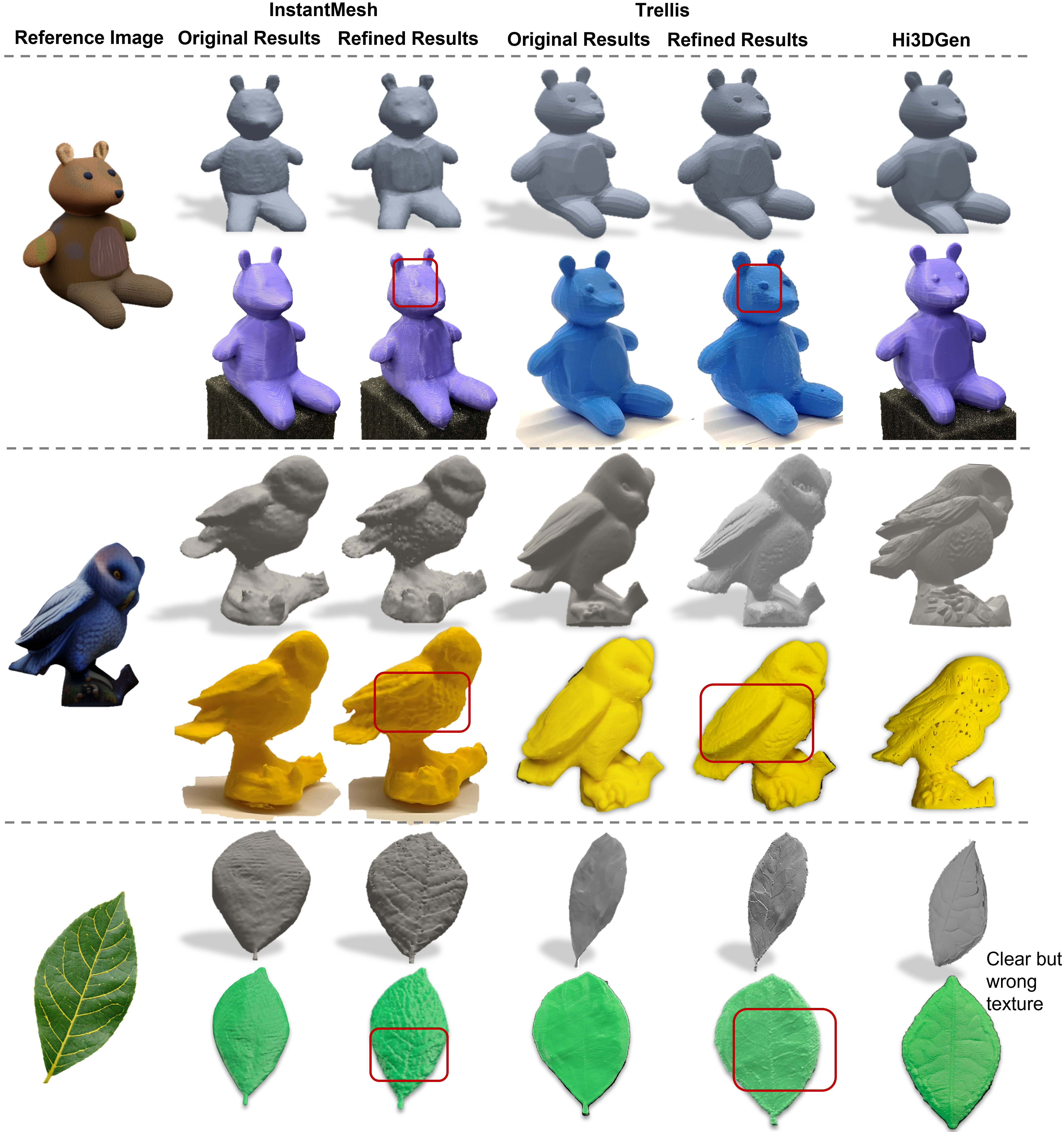}
    \caption{Physical 3D printing validation. Each object is fabricated using a single-color PLA filament. Compared with the original generated meshes, GenMF produces more recognizable surface details under monochromatic fabrication.}
    \label{fig:3dp}
\end{figure*}

Unlike prior works that focus primarily on visual realism, our evaluation explicitly targets fabrication-oriented criteria, where geometric validity and structural robustness are equally important. 

\textbf{Baselines.} We adopt recent image-to-3D approaches (InstantMesh\cite{xu2024instantmesh}, Trellis\cite{trellis}, Hi3DGen\cite{ye2025hi3dgen}) as our base model for refinement.

\textbf{Metrics.}

\textit{(1) Visual fidelity.} We use LPIPS, SSIM to measure similarity to input appearance under gray-scale. We do not report PSNR because pixel-wise intensity values are highly sensitive to illumination and are not directly comparable under different monochromatic rendering conditions.

\textit{(2) Structural robustness.} We additionally measure curvature distribution and minimum thickness. We perform finite element analysis (FEA) and report: (a) We report maximum stress and mean stress as indicators of stress concentration and overall structural robustness under the standardized simulation setting. (b) the MAE stress and the Top-5\% Stress Region IoU to report the value and distribution error of stress prediction. 

These metrics capture the inherent tension between high-frequency geometry and structural robustness.

\subsection{Qualitative Physical Fabrication Comparison}


We validate our method through real-world 3D printing. All physical prototypes in this work were fabricated using a single-nozzle Bambu Lab X1 Carbon FDM 3D printer equipped with a 0.4 mm diameter nozzle. We adopted monochromatic PLA Basic filament as the printing material. The core slicing parameters were configured as follows: a standard layer height of 0.2 mm, a sparse infill density of 12\%, tree-style support structures, and an outer-only brim to enhance first-layer adhesion. All other auxiliary printing parameters followed the factory default preset of Bambu Studio to guarantee experimental fairness and reproducibility.

As shown in Fig.~\ref{fig:3dp}, geometries produced by detail-focused methods contain fragile thin structures that are prone to failure during fabrication. Smoothing-based methods are robust but lack geometric detail. 

In contrast, our method produces stable and visually faithful results, demonstrating its practical applicability.

\subsection{Appearance Preservation under Monochromatic Fabrication}

We evaluate the ability of different methods to preserve visual appearance after monochromatic fabrication by converting texture-dependent appearance cues into geometric structures. Quantitative evaluation is conducted by rendering all reconstructed meshes under a monochromatic material and comparing the rendered images against the original textured references using LPIPS and SSIM, also with these metrics under gradient-scale(g-LPIPS and g-SSIM) to show the distribution error.

Before optimization, the outputs produced by existing 3D generation methods exhibit noticeable appearance degradation under monochromatic rendering. Since many visual details are encoded in textures rather than geometry, important structures such as surface patterns, semantic boundaries, and fine appearance cues disappear once color information is removed. Although Trellis and Hi3DGen generally preserve more geometric detail than InstantMesh due to their higher geometric resolution, substantial appearance information is still lost in monochromatic settings.

\begin{figure}
    \centering
    \includegraphics[width=0.98\linewidth]{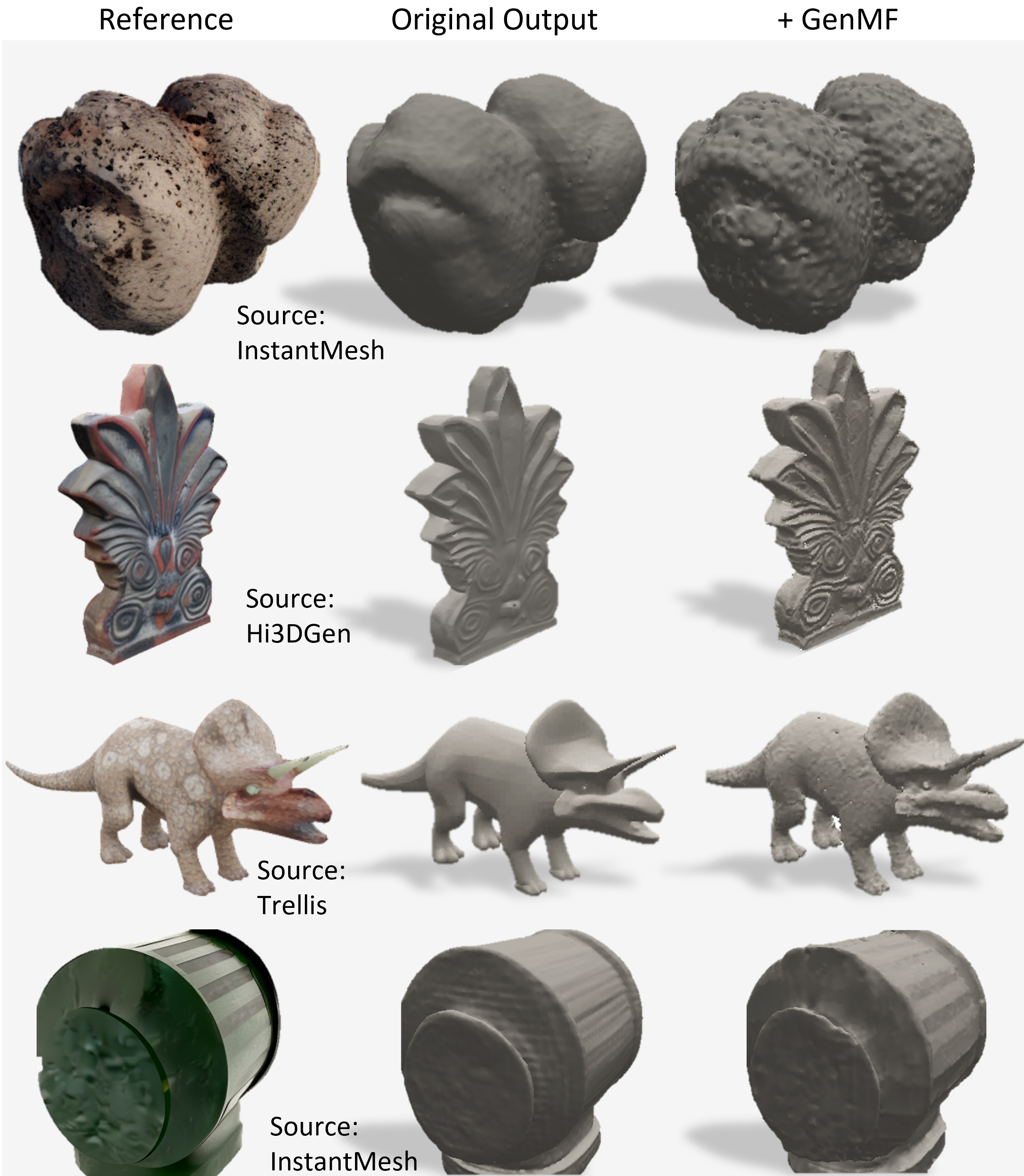}
    \caption{Appearance before and after optimization under monochromatic fabrication}
    \label{fig:resapppre}
\end{figure}

After applying GenMF, the generated geometry is refined to explicitly encode appearance cues through surface relief and shading variations. As shown in Fig.~\ref{fig:resapppre}, details that are originally represented only by texture, becoming visible in the monochromatic renderings. Instead of pursuing geometrically exact reconstruction, GenMF intentionally modifies local geometry to reproduce the perceptual appearance of the original asset through geometry-induced shading.

\begin{table}[t]
\centering
\caption{Appearance similarity between output non-texture (monochromatic) mesh and gray-scale input image.}
\label{tab:app_compare}
\begin{tabular}{lcccc}
\toprule
Metric & SSIM$\uparrow$& Lpips$\downarrow$& g-SSIM$\uparrow$& g-Lpips$\downarrow$\\
\midrule
InstantMesh &0.7862& 0.2129 & 0.7899&0.1566 \\
+ GenMF& 0.8173 & 0.2080 &0.8162&0.1455  \\
\midrule
Trellis& 0.8013 &0.1881 &0.8474&0.1150  \\
+ GenMF& 0.8243 &0.1530 &0.8923&0.0949  \\
\midrule
Hi3DGen& 0.8193 &0.1717  &0.8280&0.1034 \\
Fancy123&0.8021 &0.2029  &0.8069&0.1497  \\
\bottomrule
\end{tabular}
\end{table}

The quantitative results in Tab.~\ref{tab:app_compare} further support this observation, which are evaluated on 160 objects under 6 views per object; values are averaged over all views. Applying GenMF consistently improves LPIPS and SSIM, indicating better appearance preservation under monochromatic rendering.

\subsection{Structural Robustness Analysis}

\begin{figure}[b]
    \centering
    \includegraphics[width=0.9\linewidth]{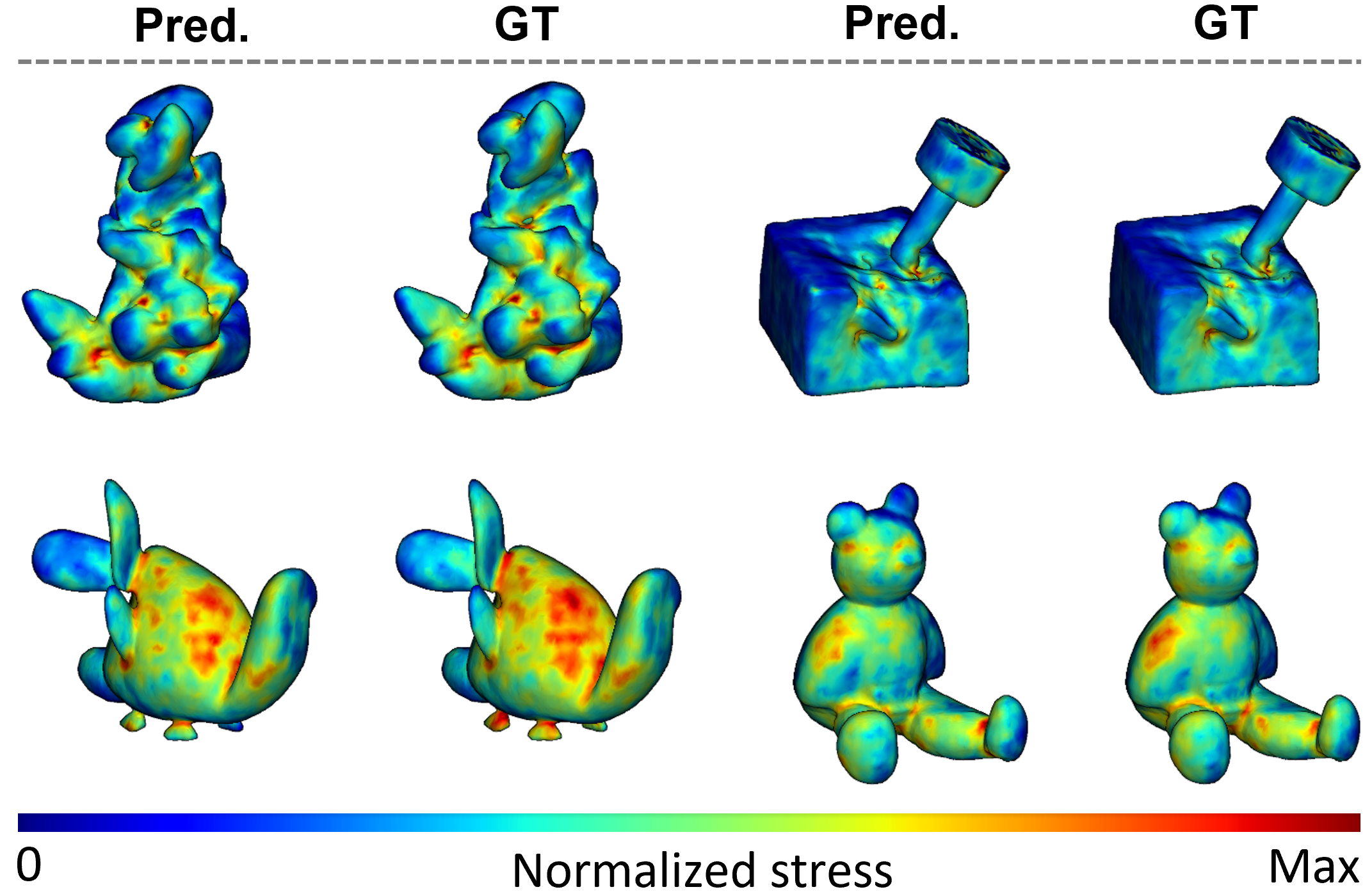}
    \caption{Validation of stress distribution prediction against FEM simulation.}
    \label{fig:stress_fig}
\end{figure}


We evaluate the structural robustness of generated meshes using finite element analysis (FEM).

For each generated mesh, we perform thermo-mechanical simulation using COMSOL under identical material properties and boundary conditions. We use PLA as the material and simulate thermal stress induced during the manufacturing process. We report the metrics introduced in Sec.~\ref{sec:sec51}.

(1) Firstly, we validate the reliability of the differentiable stress-aware objective, comparing the predicted stress fields against FEM simulations generated by COMSOL. We report mean absolute error (MAE), IoU of the top 5\% high-stress regions between predicted and simulated stress values in Tab.~\ref{tab:stress_pred}, while distribution is shown in Fig.~\ref{fig:stress_fig}.

\begin{table}[t]
\centering
\caption{Validation of stress prediction against FEM simulation.(Under our setting, the mean stress is about 0.3-0.5 MPa )}
\label{tab:stress_pred}
\begin{tabular}{ccc}
\toprule
Case&MAE$(Mpa)\downarrow$ & Top-5\% Stress Region IoU$\uparrow$ \\
\midrule
Simple Objs.&0.054&0.91\\
Complex Objs.&0.070&0.89\\
\bottomrule
\end{tabular}
\end{table}

Results show that our predictor captures major stress concentration patterns with high correlation to FEM analysis, providing reliable guidance for geometry optimization.

(2) Then we evaluate the effect of our refinement module by comparing meshes generated by each baseline method with and without our optimization under identical thermo-mechanical simulation settings.

\begin{table}[t]
\centering
\caption{Ablation on stress module. Lower is better for all metrics.(Each metric is the 0-1 ratio of the the stress without stress constraint during optimization since the stress is various from each object)}
\label{tab:fem_compare}
\begin{tabular}{lcc}
\toprule
Base Model & Max Stress$\downarrow$& Mean Stress$\downarrow$\\ 
\midrule
InstantMesh & 0.7731 &0.8357\\
Trellis& 0.6297&0.7507\\
\bottomrule
\end{tabular}

\end{table}

We visualize stress fields and deformation maps for qualitative comparison, highlighting regions of stress concentration and structural weakness.
\begin{figure*}
    \centering
    \includegraphics[width=0.8\linewidth]{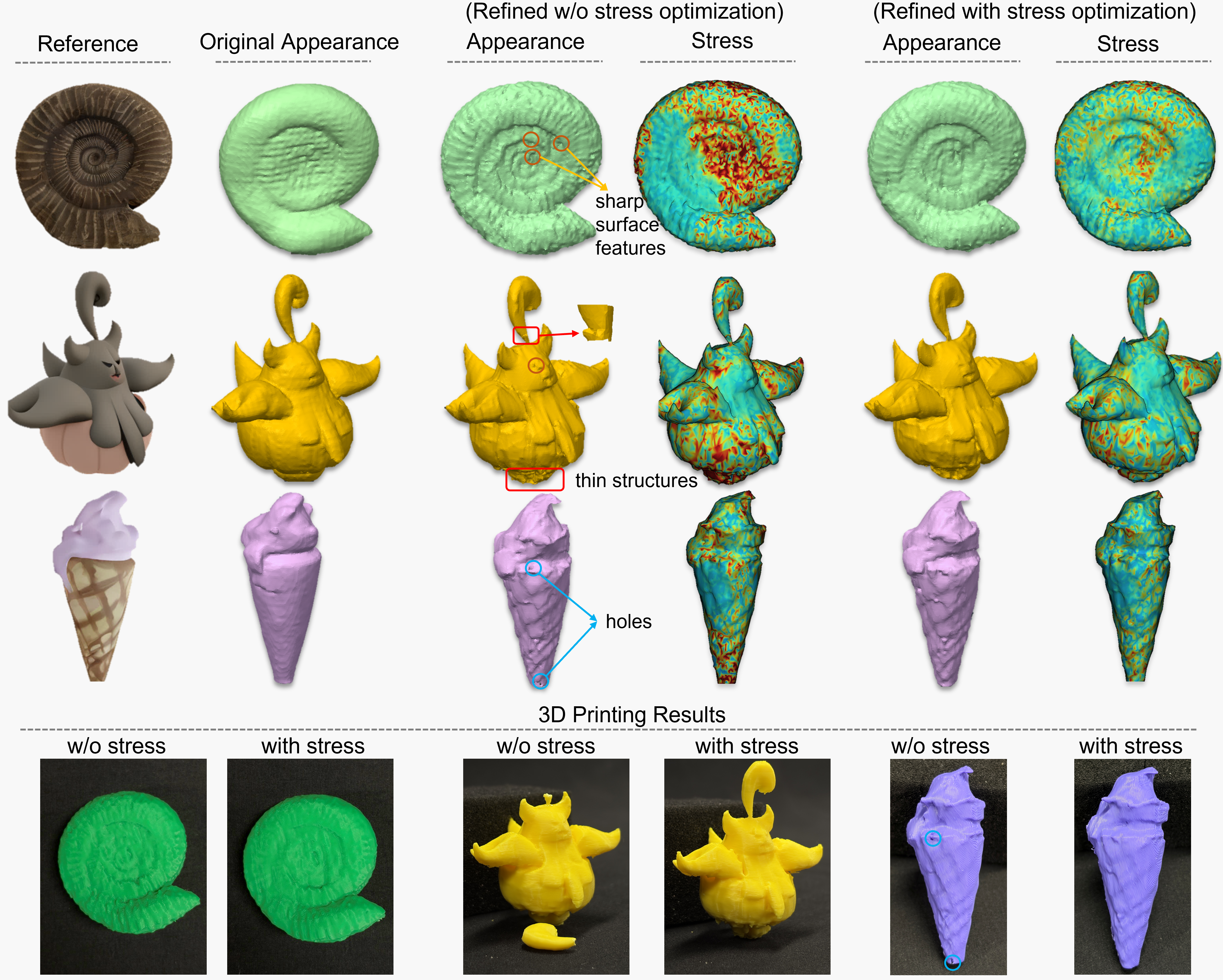}
    \caption{The refined objects and ablation on stress.}
    \label{fig:refined_fig}
\end{figure*}
This experiment demonstrates that our stress-aware optimization effectively reduces stress concentration and improves structural robustness, making the generated meshes more suitable for fabrication. Importantly, these improvements are achieved while preserving recognizable geometric appearance, demonstrating that our optimization does not merely over-smooth the generated meshes.

\subsection{Trade-off Ablation}

\begin{figure}
    \centering
    \includegraphics[width=0.98\linewidth]{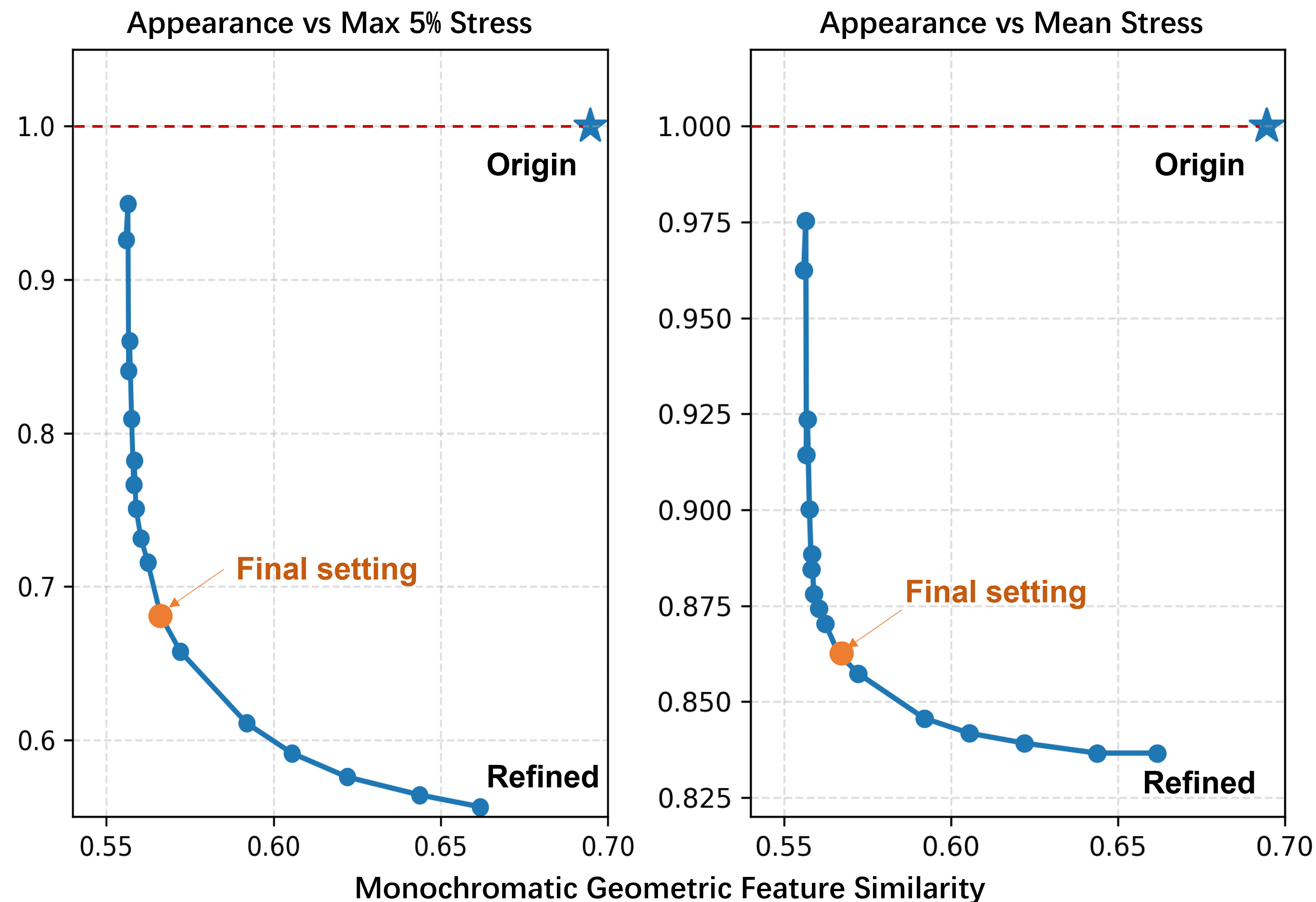}
    \caption{Trade-off ablation curve between appearance and stress condition. x-axis represents Appearance loss $\lambda_{geo}$, y-axis represents the normalized max stress / mean stress, where we set the original stress term as 1. }
    \label{fig:tradeoff}
\end{figure}

We analyze the trade-off between visual fidelity and structural robustness in our optimization framework.

We vary the weight of the stress loss ($\lambda_{stress}$) and the appearance loss ($\lambda_{gray},\lambda_{geo}$), while keeping other parameters fixed, and evaluate the resulting meshes based on Appearance quality (grayscale rendering LPIPS) and Structural robustness (maximum stress)

We plot appearance quality against structural robustness, illustrating the trade-off curve. As shown in Fig.~\ref{fig:tradeoff}.

This experiment shows that our framework provides a controllable balance between appearance and structural robustness, and can achieve better trade-offs than existing methods.




\section{Conclusion}
\label{sec:conclusion}


We present \textbf{GenMF}, an appearance-oriented geometry refinement framework for monochromatic fabrication of generated 3D assets. Instead of preserving geometry alone, GenMF converts texture-dependent appearance cues into geometry-induced shading effects, allowing visual details to remain recognizable when color information is unavailable. To avoid overly fragile geometric perturbations, GenMF further incorporates stress-aware regularization based on a learned thermal-stress predictor. Experiments show that GenMF improves monochromatic appearance preservation while reducing stress concentration under a standardized thermo-mechanical simulation setting. These results suggest that appearance-aware and fabrication-aware geometry refinement is a promising direction for bridging virtual 3D generation and physical single-material fabrication.

Our current approach is still limited in handling stress optimization over highly detailed geometric regions, where fine structures may not be fully preserved. Future work will explore higher-resolution representations and more accurate stress modeling to better capture fine-scale geometry and further improve fabrication robustness.


{
    \small
    \bibliographystyle{IEEEtran}
    \bibliography{ref}
}



\end{document}